\documentclass[review]{elsarticle}

\usepackage{lineno,hyperref}
\journal{ }

\usepackage{xcolor}

\usepackage{amsmath} 
\usepackage{graphicx}
\usepackage{amssymb} 
\usepackage{endnotes}
\usepackage{amsfonts} 
\usepackage{setspace}
\usepackage{verbatim} 
\usepackage{geometry}

\usepackage{xspace} 
\usepackage{latexsym} 
\usepackage{url}
\usepackage{amscd} 
\usepackage{multicol} 
\usepackage{algorithm}
\usepackage{rotating} 
\usepackage{array} 
\usepackage{subfigure}

\begin{document}
	
	\begin{frontmatter}

		\title{Bayesian longitudinal models for exploring European sardine fishing in the Mediterranean Sea\break {}}

		\author{Gabriel Calvo}
		\address{Departament d'Estad\'istica i Investigaci\'o Operativa, Universitat de Val\`encia, Burjassot, Valencia, Spain}
		\ead{gabriel.calvo@uv.es}
		
		\author{Carmen Armero}
		\address{Departament d'Estad\'istica i Investigaci\'o Operativa, Universitat de Val\`encia, Burjassot, Valencia, Spain}

		\author{Maria Grazia Pennino}
		\address{Instituto Espa\~{n}ol de Oceanograf\'{i}a, Centro Oceanogr\'afico de Vigo, Vigo, Pontevedra, Spain}

		\author{Luigi Spezia}
		\address{Biomathematics \& Statistics Scotland, Aberdeen, UK}
		
		\begin{abstract}
			In the Mediterranean Sea, catches are dominated by small pelagic fish, representing nearly the 49\% of the total harvest. Among them, the European sardine (Sardina pilchardus) is one of the most commercially important species showing high over-exploitation rates in recent last years.  In this study we analysed the European sardine landings in the Mediterranean Sea from 1970 to 2014. We made use of Bayesian longitudinal linear mixed models in order to assess differences in the temporal evolution of fishing between and within countries. Furthermore, we modelled the subsequent joint evolution of artisanal and industrial fisheries.  Overall results confirmed that Mediterranean fishery time series were highly diverse along their dynamics and this heterogeneity was persistent throughout the time. In addition, results highlighted a positive relationship between the two types of fishing.
		\end{abstract}
		
		
	\end{frontmatter}
%	\linenumbers
	
	\section{Introduction}
	\label{sec1}
	Small pelagic fish species have been proven to be key elements of the Mediterranean pelagic ecosystem due to their high bulk biomass at the mid-trophic level, which provides an important energy connection between the lower and upper trophic levels \cite{albo2015feeding}. Indeed, historically they have shown to be essential in the coupling between the pelagic and the demersal environment, as they are prey for pelagic predators such as tuna \cite{navarro2017feeding}, cetaceans \cite{gomez2011overfishing}, pelagic birds \cite{navarro2009seasonal} and demersal predators such as hakes \cite{mellon2017trophic}. Fluctuations in populations of small pelagics can therefore have serious ecological and socio-economic consequences. For this reason, understanding their underlying mechanisms is extremely important \cite{queiros2019food}. The population dynamics of these species could be strongly influenced by natural environmental fluctuations (bottom-up control) and mortality (top-down control) \cite{checkley2017climate}. In particular, overfishing has also been identified as a key factor in the collapse of several populations of small pelagics, often in combination with environmental fluctuations and under a global climate change context \cite{brochier2013climate}.
	
	Small pelagic fishes usually live in dense shoals, making gear such as mid-water pelagic trawls and purse seines particularly efficient for their capture. Indeed, in the Mediterranean sea, catches are dominated by small pelagics, representing nearly 49\% of the harvest \cite{saraux2014spatial}. Among them, the European sardine ({\sl Sardina pilchardus}, Walbaum, 1792) is one of the most commercial species, and is the third species, after the European hake ({\sl Merluccius merluccius}) and the red mullet ({\sl Mullus barbatus}). Furthermore it has shown the highest over-exploitation rates in the last years \cite{coll2008food}. 
	
	At regional level, strong declines in sardine landings have been reported in different Mediterranean areas \cite{coll2019blame}. Causes of these recent  declines have been related to high fishing impact, competition between pelagic organisms or ecosystem effects \cite{brosset2017spatio, quattrocchi2017environmental, brigolin2018using}.
	
	However, no study has been performed to assess sardine landings dynamics at large- e.g. basin-scale. Although working at fine scale and regional level (e.g. Mediterranean eco-regions or subareas) has several advantages, large-scale analysis are essential to provide a broader perspective of non-localized drivers of fisheries where sardine fish stocks are shared among countries, and to reduce the bias of landings misallocation \cite{stergiou2016trends}.
	Mediterranean fisheries are highly diverse and geographically varied, not only because of the existence of different marine environments, but also because of different socio-economic situations, and fisheries status \cite{pennino2017analysis}. The heterogeneity between countries can be investigated through study of the behavior of fishery exploitation across different Mediterranean countries. 
	
	Within this context, in this study we analyse the European sardine landings in the Mediterranean Sea from 1970 to 2014. We make use of Bayesian longitudinal linear mixed models in order to assess  differences in the temporal evolution of fishing between and within countries. Furthermore,  we model  the subsequent joint evolution of artisanal and industrial fisheries which are defined in terms of small-scale and large-scale commercial fisheries, respectively \cite{zeller2015reconstructing}.

	\section{Material and methods}
	\label{sec2}
	
	\subsection{European sardine fishing}
	Landing data (tonnes) of the European sardine caught by Mediterranean countries were extracted from  \textsf{Sea Around Us} (www.seaaroundus.org) from 1970 to 2014.
	Countries participating in the study are Albania, Algeria, Bosnia and Herzegovina, Croatia, France, Greece, Italy, Montenegro, Morocco, Slovenia, Spain and Turkey. 
	Data from countries recognized as sovereign   from    2010 by the international community (Bosnia and Herzegovina, Croatia, Montenegro, and Slovenia) were imputed from 1950-2010 \cite{zeller2015reconstructing} based on information from Exclusive Economic Zones (EEZ), which was linked to these countries after reconstructing data by means of several sources.
	Despite the fact that the original data set also included subsistence and recreational landings,  data from artisanal and industrial fisheries were only used in this study because it is only focused   on commercial fisheries.
	Finally,  a logarithmic transformation was applied to the landing data in order to approach the normality assumption. 
	
	\subsection{Modelling European sardine landings}
	The term longitudinal data is generally used to indicate that each subject of the sample is measured repeatedly on the same outcome at several points in time. The main objective in this type of study is to detect and analyse general patterns of the temporal evolution of the target population  as well as relevant individual characteristics.  For this reason, each Mediterranean country was considered as an annually observed individual. Annual European sardine landings formed a time series (1970 to 2014) for each country.

	The assumption that all the information about the response variable is generated by a function common to all individuals of the target population is not realistic in longitudinal designs. Ignoring individual heterogeneity among individuals, in this case the countries,  could lead to inconsistent and inefficient estimates of the parameters of interest \cite{pinheiro2000}. As a result, our longitudinal models include not only  general population information and measurement error terms \cite{diggle2002analysis} but also  random effects to account for specific country characteristics.  Statistical estimation is carried out within the Bayesian inferential framework. Consequently, we assume a conception of the probability that allows us to assign probability distributions to all unknown quantities. The posterior distribution for the relevant parameters, hyperparameters and random effects contains all available information of the problem. This distribution was approximated in all modelling via the JAGS software \cite{plummer2003jags} through Markov chain Monte Carlo (MCMC) sampling.
	
	The  content of this paper involved the construction and assessment of: (1) a longitudinal mixed model to assess the temporal dynamic of the total European sardine landings by country, and (2) a joint longitudinal mixed model to assess the temporal dynamic of the sardine landings by country with regard to the type of fisheries, industrial and artisanal.

	%Longitudinal models for assessing the temporal evolution of the variable of interest, the annual European sardine landings by country, included the general population information, random effects which account for specific country characteristics, and measurement error term \cite{diggle2002analysis}. 
	%The most popular and paradigmatic longitudinal model is the lineal mixed-effects model \cite{pinheiro2000}. 

	\subsubsection{A longitudinal model for the total European sardine landings}
	
	Let $Y_{it}$ be the logarithm of total tonnage of sardines  caught in   country $i$ during year $t$, $t=0,\dots,T-1$. Calendar time is the natural time scale of the study and time zero  ($t=0$) is the first year of the study (i.e. $1970$).
	
	We considered a Bayesian linear mixed-effects models  that expresses  the conditional distribution of the random variable $Y_{it}$   in terms of the normal distribution   $\text{N}(\beta_0 + b_{0i} + b_{1i} t, \sigma^2)$ whose   mean includes a common intercept  ($\beta_0$) and two specific  terms associated with the individual characteristics of the countries:   a random effect  associated with the intercept ($b_{0i}$) and a random effect for   the  slope $(b_{i1})$. This model is homoscedastic: the conditional variance  $\sigma^2$ is the same for all countries and remains constant over time.  Random effects are assumed conditionally independent and normally distributed. In particular,   $\boldsymbol{b}_i= (b_{0i}, b_{1i})$,  where $f(b_{0i}|\sigma^2_0 )= \text{N}(0, \sigma^2_0)$ and $f(b_{1i}|\sigma^2_1)= \text{N}(0, \sigma^2_1)$.
	
	We propose a Bayesian inferential process which gives the maximum  prominence to  the data. We assume a non-informative prior scenario, with prior independence,  and select a normal distribution with large standard deviation for the common regression coefficient $\pi(\beta_0)=\text{N}(0,10^2)$, and uniform distributions for all standard deviation terms $\pi(\sigma)=\pi(\sigma_0)=\pi(\sigma_{1})=\text{U}(0,10)$.

	\subsubsection{A joint longitudinal model for the industrial and artisanal European sardine landings}
	We also propose a Bayesian joint longitudinal model for analysing the dynamic behaviour and relationships of the European sardine landings caught by artisanal or industrial fisheries.   Let   $Y_{it}^{(I)}$ and $Y_{it}^{(A)}$ be the landing of the European sardine caught by artisanal and industrial  fishery in the country $i$  during year $t$, respectively. 
	
	We  assume  a Bayesian share-parameter model
	to jointly model both processes \cite{Verbeke2009, armero2018} that uses the random effects to generate   an association structure  between both longitudinal measures. To this effect, the   joint  distribution of $Y_{it}^{(I)}$, $Y_{it}^{(A)}$, parameters and hyperparameters $\boldsymbol{\theta}$, and random effects for country $i$  can be expressed as:
	\begin{equation*}
		\begin{split}
			f(y_{it}^{(I)}, y_{it}^{(A)}, \boldsymbol{\theta}, \boldsymbol{b}_i)&= f(y_{it}^{(I)}, y_{it}^{(A)}| \boldsymbol{\theta}, \boldsymbol{b}_i)f(\boldsymbol{b}_i|\boldsymbol{\theta})\pi(\boldsymbol{\theta}) \\
			&= f(y_{it}^{(I)}| \boldsymbol{\theta}, \boldsymbol{b}_i^{(I)})
			f(y_{it}^{(A)}| \boldsymbol{\theta}, \boldsymbol{b}_i^{(A)})
			f(\boldsymbol{b}_i|\boldsymbol{\theta})\pi(\boldsymbol{\theta}),
		\end{split}
	\end{equation*}
	\noindent where
	\begin{equation}\label{eq:industrial_artisanal_formulation}
		\begin{split}
			f(y_{it}^{(I)}| \boldsymbol{\theta}, \boldsymbol{b}_i^{(I)}) &=  \text{N}(\beta_0^{(I)} + b_{0i}^{(I)} + b_{1i}^{(I)} t,\sigma^2),\\
			f(y_{it}^{(A)}| \boldsymbol{\theta}, \boldsymbol{b}_i^{(A)})&=  \text{N}(\beta_0^{(A)} + b_{0i}^{(A)} + b_{1i}^{(A)} t,  \sigma^2).
		\end{split}
	\end{equation}
	The random effects vector for country $i$ can be divided in two subvectors, $\boldsymbol{b}_i = (\boldsymbol{b}_i^{(I)}, \boldsymbol{b}_i^{(A)}) $ corresponding to the industrial and artisanal fishing, respectively. In addition, we impose a structure of association   between the random effects associated with the industrial  and artisanal fishing,  $f(b_{0i}^{(I)},b_{0i}^{(A)}|\Sigma_0)=\text{N}(0,\Sigma_0) $ and $f(b_{1i}^{(I)},b_{1i}^{(A)}|\Sigma_1)=\text{N}(0,\Sigma_1) $, with variance - covariance matrices given by

	\begin{equation*}
		\Sigma_0 = \begin{pmatrix}
			{\sigma_0^{(I)}}^2 & \rho_0 \sigma_0^{(I)} \sigma_0^{(A)}\\ 
			\rho_0 \sigma_0^{(I)} \sigma_0^{(A)} & {\sigma_0^{(A)}}^2
		\end{pmatrix},\ \ \ \Sigma_1 = \begin{pmatrix}
			{\sigma_1^{(I)}}^2 & \rho_1 \sigma_1^{(I)} \sigma_1^{(A)}\\ 
			\rho_1 \sigma_1^{(I)} \sigma_1^{(A)} & {\sigma_1^{(A)}}^2
		\end{pmatrix}.
	\end{equation*}
	The inferential process is developed using the same prior distributions set before for the total amount of European sardine landing caught, with the addition of a uniform prior for the correlations:  $\pi(\rho_0)=\pi(\rho_1)=\text{U}(-1,1)$.
	%%%%%%%%%%%%%%%%%%%%%%%%%%%%%%%%%%%%%%%%%%%%%%%%%%%%%%%%%%%%%%%%%%%%%%%%%%%%%%%%%%%%%%%%%%%%%%%%%%%%%%%%%%%%%%%%%%%%%%

	\section{Results}
	Top graph of Figure 1 shows the spaghetti plot (data for each country connected across
	time [Hedeker and Gibbons, 2006]) of the total amount of sardines caught, on a logarithmic
	scale, from 1970 to 2014 in all the  countries  in the study. Middle and bottom graphs of Figure 1 are the subsequent plots for industrial and artisanal
	fishing, respectively. Note that not all countries have both types of fisheries (Bosnia and Herzegovina has no registered industrial fisheries and Albania and Spain have no artisanal fisheries). %\vspace{-4cm}

	%Figure \ref{fig:spaghetti_plots} on the top shows the spaghetti plot (data for each country connected across time \cite{hedeker2006longitudinal}) of the total amount of European sardine caught, in logarithmic scale, from 1970 to 2014 in all the Mediterranean countries included in the study. Figure \ref{fig:spaghetti_plots} in the middle and in the bottom are the subsequent plots for industrial and artisanal fishing, respectively.

	%Tonnes of catched sardines can only take positive values. We have worked in the
	%logarithmic scale and not with the raw scale of the data in order to approach to normality
	%in the statistical analysis. It is also important to realize the relevance of the values in
	%that scale. For example, going from a difference of 9 to 10 tonnes on the logarithmic scale
	%means moving from 8103 to 22026 tonnes approximately.

	Overall, the graphs show  a very similar behaviour:  the amount of fish caught at the beginning of the study is very different among the countries which, in most of the cases, exhibit an evident linear trend. Artisanal fisheries generally catch less sardines than industrial fisheries, which seem to experience more stable dynamic patterns. Bosnia and Herzegovina is the country that fishes the least sardines. It only reports industrial fishing but has experienced a notable expansion in the most recent years. Turkey also increased fishing significantly between the 70s and the 90s, and it is currently,  along with  Algeria, Croatia, Greece and Spain, has one of the largest sardine fisheries. France and Slovenia have lowered their catches in recent years.

	\subsection{A longitudinal model for the total European sardine landings}
	
	Table \ref{tab:RITM} shows the approximate posterior mean, standard deviation and 0.025 and 0.975-quantiles for the  parameters  of the model. Random effects are very relevant in the estimated model. The posterior mean associated with the standard deviation $\sigma_0$ of   the random intercept is relatively high, indicating a great variability  in the amount of fishing in the countries at the beginning of the study.   The posterior mean of the standard  deviation of the random slopes, $\sigma_1$, has a small value but results in large variations when we leave the logarithmic scale and the study time increases. These results highlight the fact that   each country shows a particular pattern in the evolution of the European sardine landings dynamic, especially in the early years of the study.
	
	\begin{figure}[H]
		\centering
		\includegraphics[width=120mm]{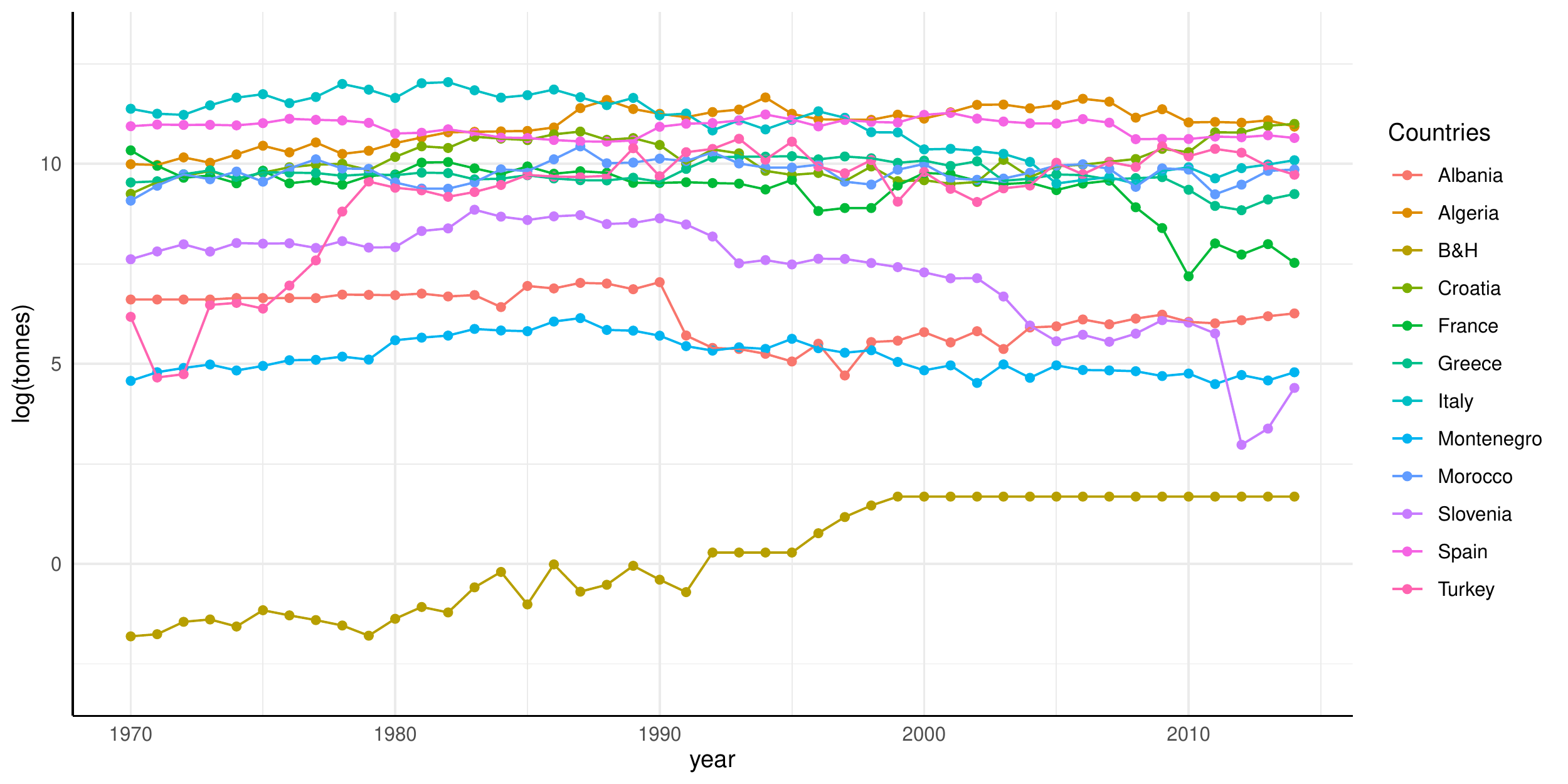}\\ %\vspace*{-5.3cm}
		\includegraphics[width=120mm]{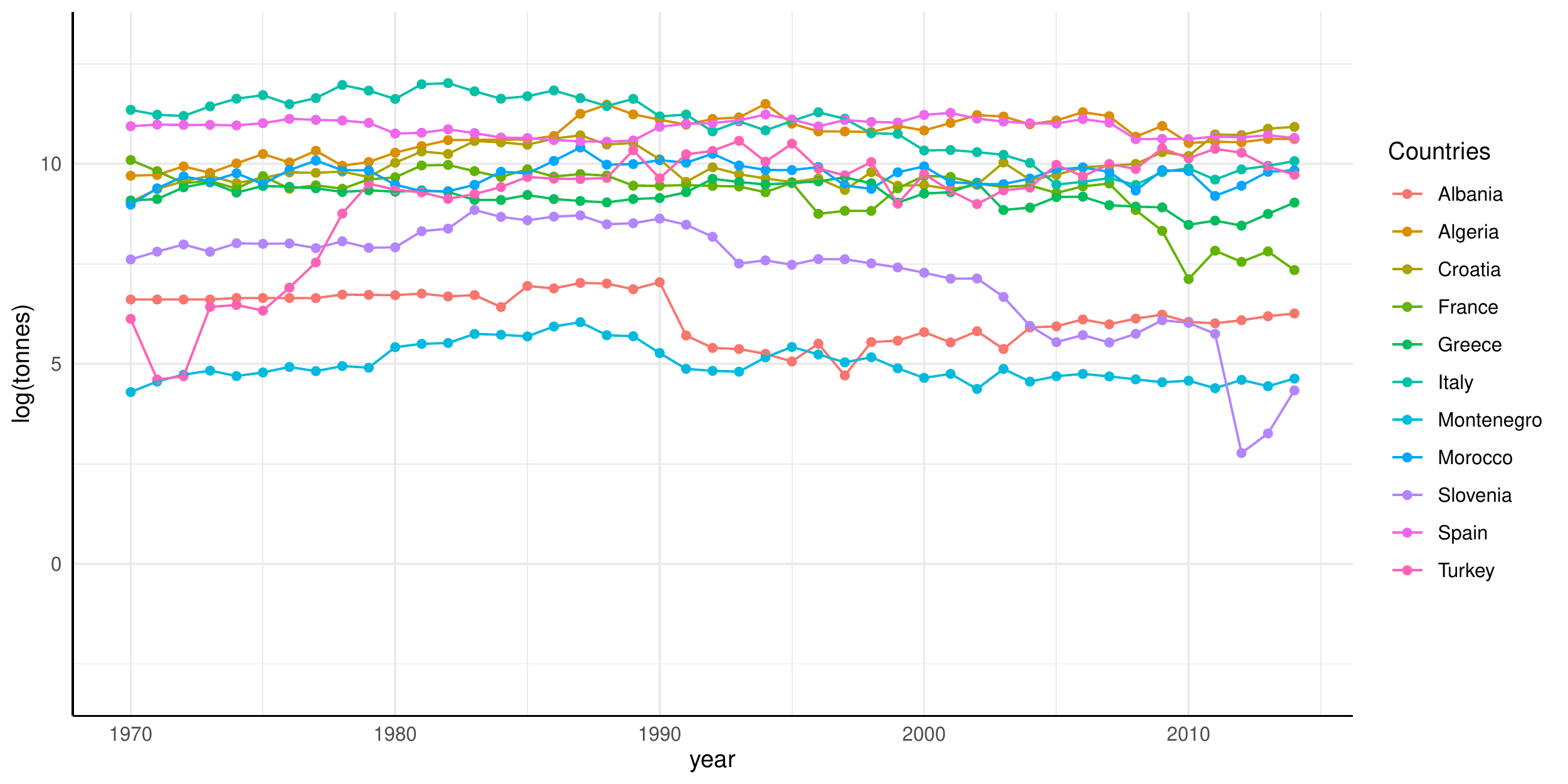}\\%\vspace*{-5.3cm}
		\includegraphics[width=120mm]{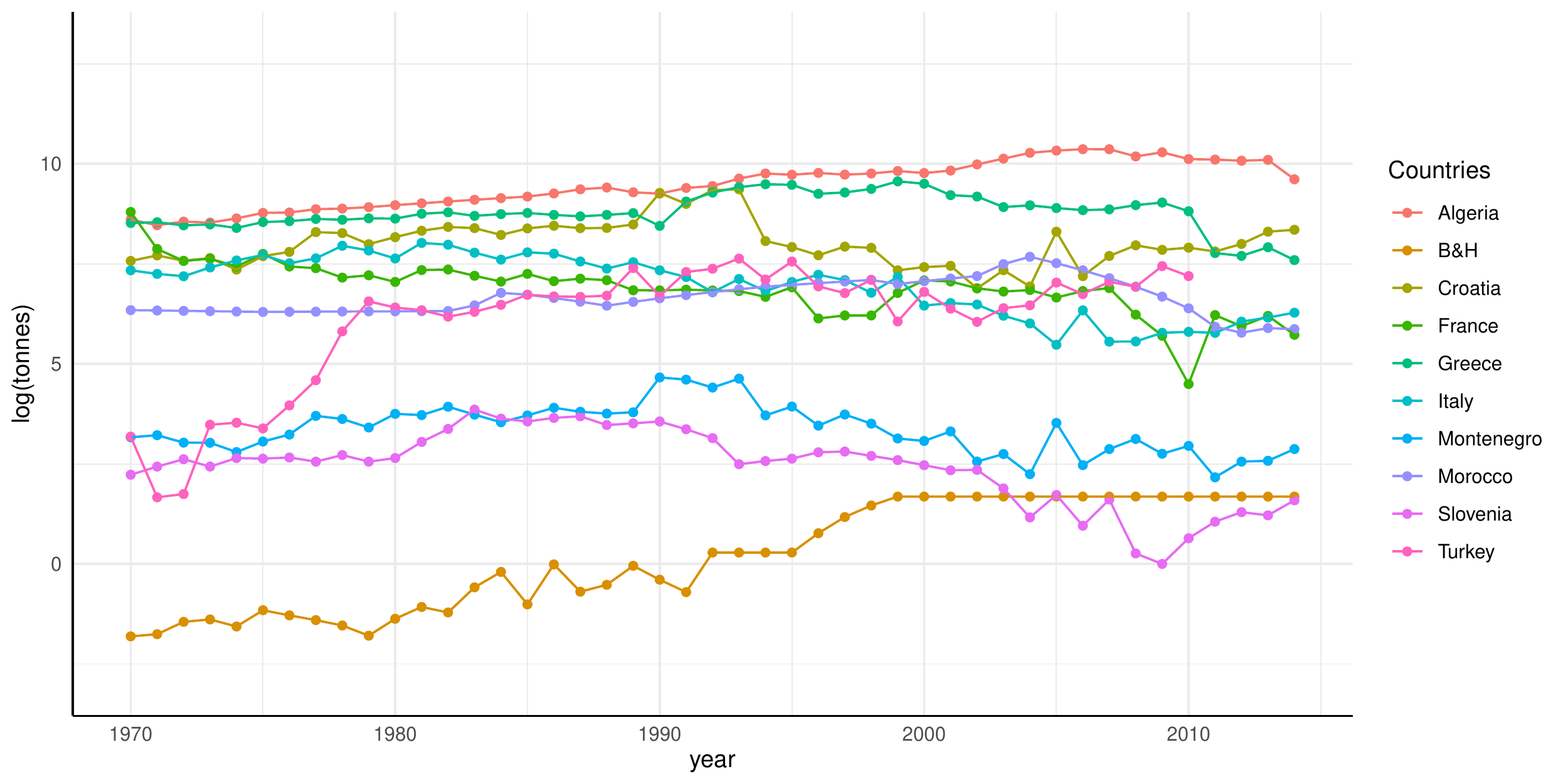}\\ %\vspace*{-1cm}
		\caption{Annual log tons of European sardine ({\sl Sardina pilchardus}) per country, from 1970 to 2014. From top to bottom: total, industrial, and artisanal fishing.} \label{fig:spaghetti_plots}
	\end{figure}
	
	%Table \ref{tab:RITM} shows the approximate posterior mean, standard deviation and 0.025 and 0.975-quantiles for the  parameters and hyperparameters of the model. 

	%{\rojo which means that ....}{\gabriel every country has a particular and unique dynamic. }{\rojo what imply that standard  deviation of the random slopes, $\sigma_1$, has a small value?} {\gabriel  although it looks small, in logarithmic scale it isn't.}

	\begin{table}[H]
		\centering
		\caption{Posterior summaries  for the parameters of the longitudinal linear mixed model for the total amount of European sardine ({\sl Sardina pilchardus}) caught.}\label{tab:RITM}
		\vspace{0.25cm}
		\begin{tabular}{ccccc}
			\noalign{\hrule height 1pt}
			\multicolumn{1}{l}{}           & \multicolumn{4}{c}{Total fishing}     \\ \cline{2-5}
			\multicolumn{1}{l}{} & \multicolumn{1}{c}{mean}    & \multicolumn{1}{c}{sd} & $q_{0.025}$ & $q_{0.975}$ \\ \noalign{\hrule height 1pt}
			\multicolumn{1}{c}{$\beta_0$}  & \multicolumn{1}{r}{$8.098$} & \multicolumn{1}{r}{$1.260$} & \multicolumn{1}{r}{$5.619$} & \multicolumn{1}{r}{$10.446$} \\
			\multicolumn{1}{c}{$\sigma$}     & \multicolumn{1}{r}{$0.541$} & \multicolumn{1}{r}{$0.017$} & \multicolumn{1}{r}{$0.510$}& \multicolumn{1}{r}{$0.574$}\\
			\multicolumn{1}{c}{$\sigma_0$}   & \multicolumn{1}{r}{$4.234$} & \multicolumn{1}{r}{$1.028$} & \multicolumn{1}{r}{$2.723$}& \multicolumn{1}{r}{$6.598$}\\
			\multicolumn{1}{c}{$\sigma_1$}   & \multicolumn{1}{r}{$0.054$} & \multicolumn{1}{r}{$0.014$}& \multicolumn{1}{r}{$0.035$}& \multicolumn{1}{r}{$0.089$}\\ \noalign{\hrule height 1pt}
		\end{tabular}
	\end{table}

\begin{figure}[H]
		\centering
		\subfigure[Random intercepts. ]{\includegraphics[width=75mm]{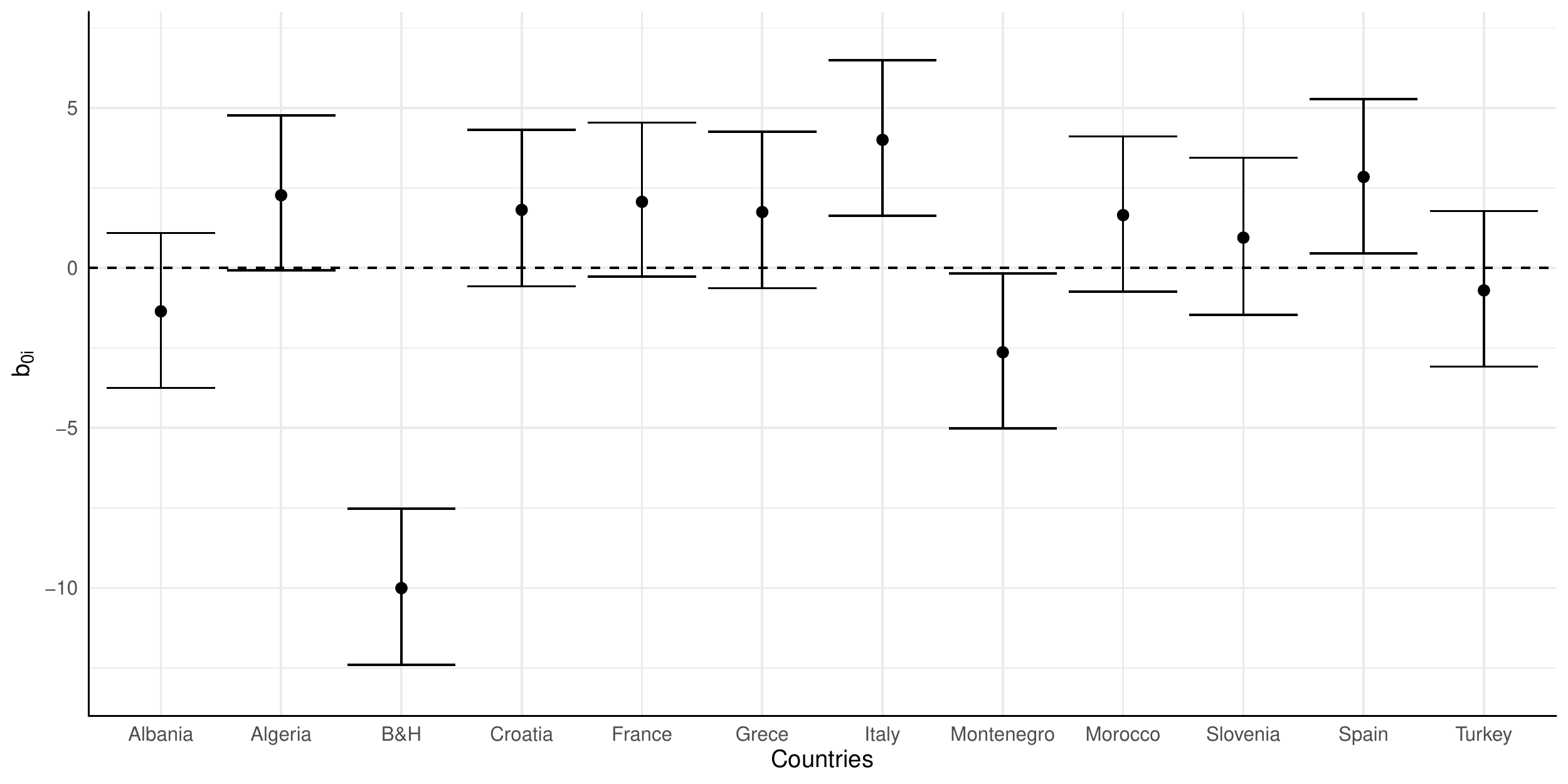}}
		\subfigure[Random slopes.]{\includegraphics[width=75mm]{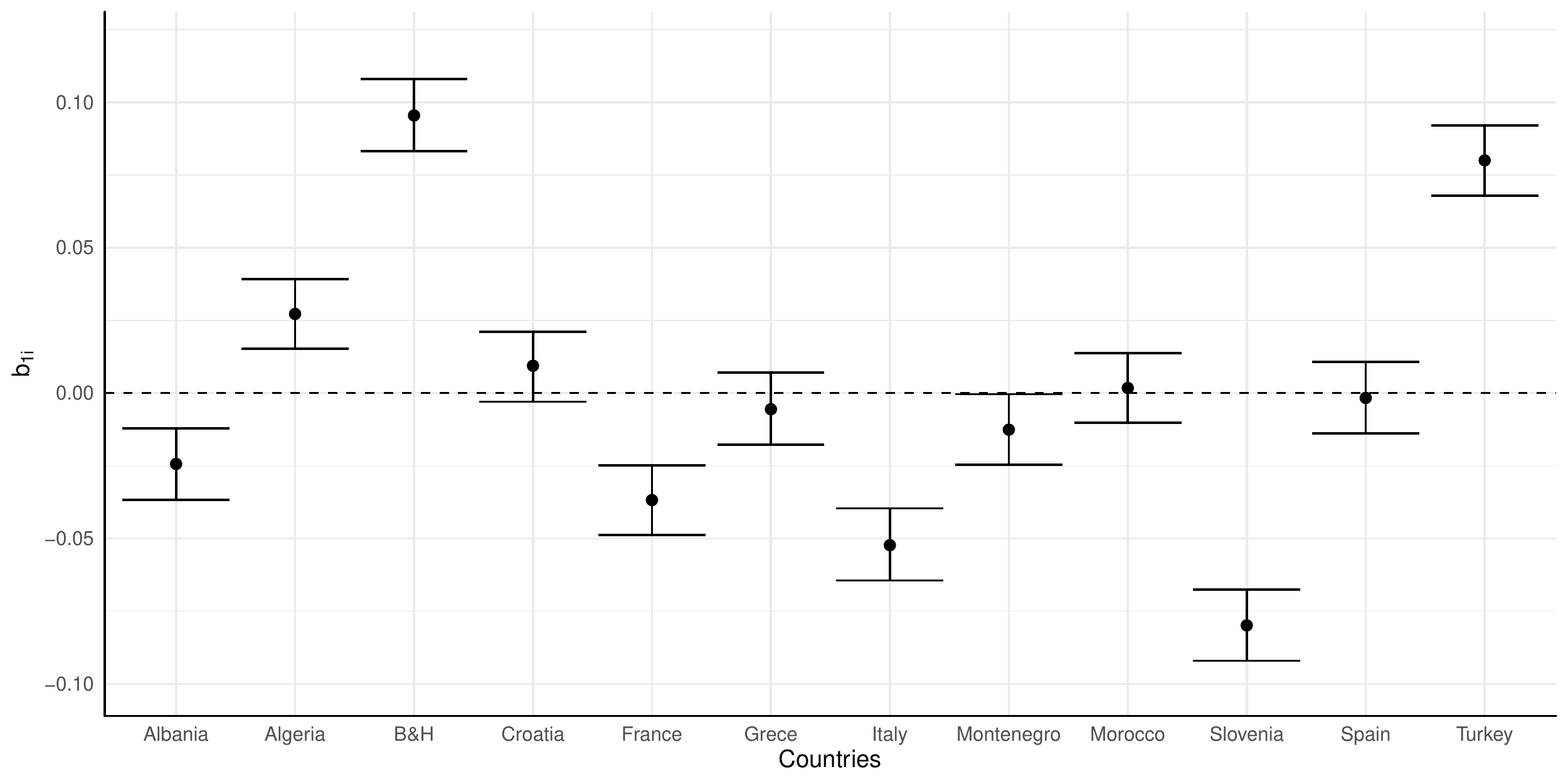}}
		\caption{Posterior 95 \% credible intervals for the random intercepts and the slope effects by country for the total European sardine ({\sl Sardina pilchardus}) landings model.}
		\label{fig:random_effects_linear_mixed_model}
	\end{figure}

	Figure \ref{fig:random_effects_linear_mixed_model} shows an approximate posterior 95 \% credible interval for the  random intercept and slope effects associated with each country. In the case of the random intercepts, we can see that Spain and Italy have positive intervals which means that both countries are above the starting point average (Figure  \ref{fig:random_effects_linear_mixed_model}). On the other hand, Bosnia and Herzegovina and Montenegro (with imputed data before their international recognition in 2010) are the only countries whose credible interval is entirely negative so their starting point are the lowest. The rest of the countries have values quite close to the general mean at the beginning of the study (Figure  \ref{fig:random_effects_linear_mixed_model}).
	
	Posterior credible intervals for the random slope associated with each country make evident the  temporal evolution in the European sardine fisheries of Mediterranean countries. Countries with 95 \% credible intervals completely positive indicate an increasing catch while completely negative intervals indicate the opposite pattern. Otherwise, the dynamic is considered stable. Accordingly, countries such as Algeria, Bosnia and Herzegovina and Turkey exhibit positive trends. On the other hand, Albania, France, Italy, Montenegro and Slovenia show decreasing patterns and Croatia, Greece, Morocco and Spain quite stable.
	
	\subsection{A joint longitudinal model for the industrial and artisanal European sardine landings}
	
	Table \ref{tab:joint_model2} summarizes the approximate mean, standard deviation and 0.025 and 0.975 quantiles for the   posterior distribution for the parameters of the joint model.  The  mean of the common intercept associated with  industrial fishing  is greater than the subsequent for artisanal and therefore  the industrial activity  was bigger than the artisanal one in the 70s, the beginning of the period of the study. Random effects are again relevant, with similar results to those obtained for the  longitudinal model that accounted for the total landings. However, standard deviation parameters for random effects in the artisanal fishing variable are higher than those which are associated with the industrial fishing variable, especially  the random effects associated with the intercept. This indicates a greater disparity between countries in artisanal fisheries than in industrial fisheries in the early years of the study.
	
	Results for the  correlation parameters  provide interesting information: the posterior mean for $\rho_0$ and $\rho_1$ are $0.673$ and $0.900$, respectively. That points to a positive correlation between industrial and artisanal country random effects, especially for random slopes that show stronger correlation. This fact indicates  a relevant synergy relationship between both types of fishing.   The credibility interval of $\rho_0$, (-0.029, 0.946), is  not very informative,  but provides evidence for a positive relationship between both intercept random effects.

	\begin{table}[H]
		\centering
		\caption{Posterior summaries for the joint model for the amount of industrial (I) and artisanal (A) European sardine ({\sl Sardine pilchardus}) fishing}\label{tab:joint_model2}
		\vspace{0.25cm}
		\begin{tabular}{crrrr}
			\noalign{\hrule height 1pt}
			\multicolumn{1}{c}{}                & \multicolumn{4}{c}{Artisanal and industrial fishing}                             \\ \cline{2-5}
			\multicolumn{1}{c}{}      & \multicolumn{1}{c}{mean}     & \multicolumn{1}{c}{sd} & \multicolumn{1}{c}{$q_{0.025}$ } &
			\multicolumn{1}{c}{$q_{0.975}$ }\\ \noalign{\hrule height 1pt}
			\multicolumn{1}{c}{$\beta_0^{(I)}$} & \multicolumn{1}{r}{$8.731$}  & $0.875$        & $6.893$ &  $10.497$      \\
			\multicolumn{1}{c}{$\beta_0^{(A)}$} & \multicolumn{1}{r}{$5.651$}  & $1.208$        & $3.299$ &  $8.201$      \\
			\multicolumn{1}{c}{$\sigma$}      & \multicolumn{1}{r}{$0.565$}  & $0.014$    & $0.538$ &  $0.592$          \\
			\multicolumn{1}{c}{$\sigma_0^{(I)}$}    & \multicolumn{1}{r}{$2.648$} & $0.786$    & $1.563$ &      $4.536$      \\
			\multicolumn{1}{c}{$\sigma_0^{(A)}$}    & \multicolumn{1}{r}{$3.823$} & $1.014$    & $2.388$ &      $6.307$      \\
			\multicolumn{1}{c}{$\sigma_1^{(I)}$}    & \multicolumn{1}{r}{$0.051$}  & $0.013$ & $0.032$ & $0.081$ \\
			\multicolumn{1}{c}{$\sigma_1^{(A)}$}    & \multicolumn{1}{r}{$0.052$}  & $0.012$ & $0.034$ & $0.081$ \\
			\multicolumn{1}{c}{$\rho_0$}      & \multicolumn{1}{r}{$0.673$}  & $0.258$   & $-0.029$ &   $0.946$          \\
			\multicolumn{1}{c}{$\rho_1$}      & \multicolumn{1}{r}{$0.900$}  & $0.097$      & $0.629$ &   $0.988$       \\\noalign{\hrule height 1pt}
		\end{tabular}
	\end{table}

\begin{figure}[H]
		\centering
		\subfigure[Random intercepts. ]{\includegraphics[width=75mm]{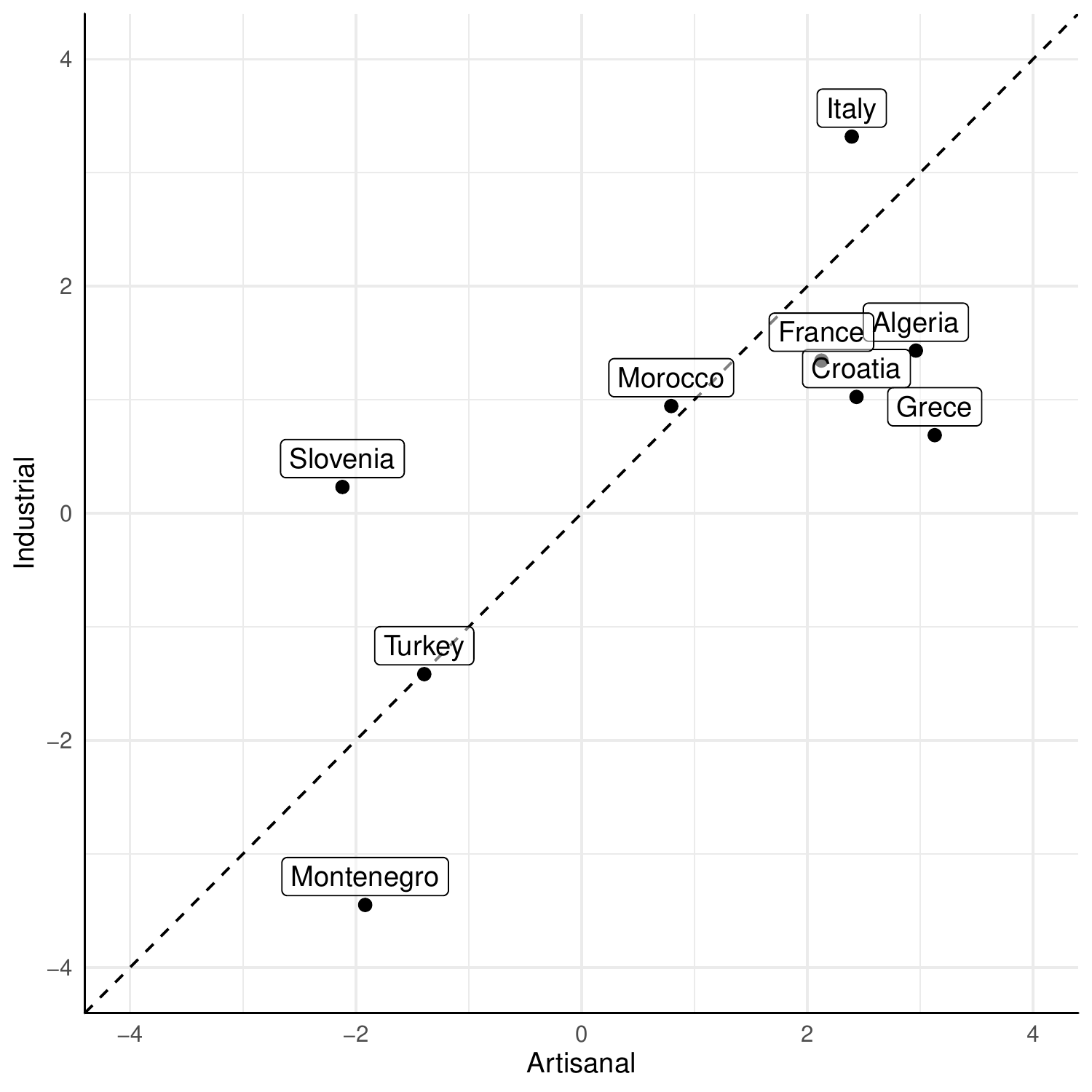}}
		\subfigure[Random slopes.]{\includegraphics[width=75mm]{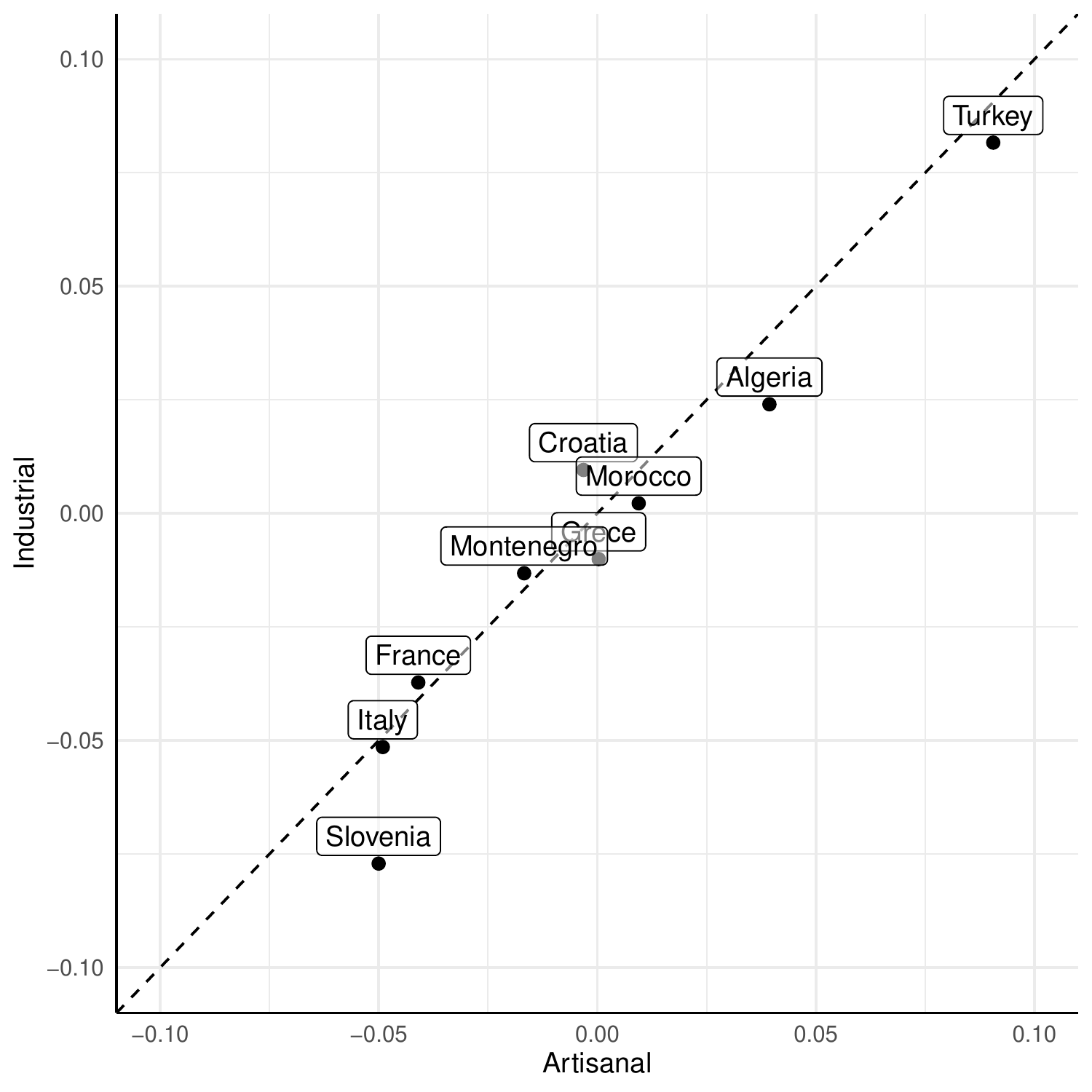}}	
		\caption{Industrial vs. Artisanal fishing. Approximate posterior mean of the random effects parameters from the joint model: (a) Random intercepts and (b) random trends.}
		\label{fig:mean_random_effects_joint_dependent_effects}
	\end{figure}

	Figure \ref{fig:mean_random_effects_joint_dependent_effects} shows the posterior  mean of the artisanal and industrial random effects (Figure 3a and 3b for the intercepts and the slopes, respectively) of the  countries that have   both types of fishing: Algeria, Croatia, France, Greece, Italy, Montenegro, Morocco, Slovenia and Turkey. In both figures, the dashed line is the identity function. As a result,  a country   above the bisector   means that its random effect associated with the industrial fishing  is greater than the one for the artisanal fishing. 
	
	Figure \ref{fig:mean_random_effects_joint_dependent_effects}a shows a positive but weak trend. Slovenia is the country that  disagrees most with the general trend of the other countries. On the other hand, the starting points of Morocco and Turkey have identical values for both types of fishing.  Figure \ref{fig:mean_random_effects_joint_dependent_effects}b shows a strong positive association between the random slopes  associated with each country for the two types of fishing considered ($b_{1i}^{(I)}$ and $b_{1i}^{(A)}$ in equation (\ref{eq:industrial_artisanal_formulation})). This  graph perfectly illustrates the dynamic of positive synergy between both types of fishing in the countries of the study. Again, Slovenia's behaviour is somewhat of an outlier. We now observe that its specificity associated with industrial fisheries is a little lower than expected in relation to its artisanal component. This situation might be due to its special behaviour at the beginning of the study.

	%Figure \ref{fig:random_effects_joint} shows an approximate  95\% credible intervals for the random intercept and slope of the joint model that accounts for the special characteristics of the countries with regard the industrial fishing. Posterior 95\% credible intervals for the random intercept associated to  Algeria, Croatia, Italy and  Spain are  completely positive. On the other hand, Bosnia and Herzegovina, and Montenegro have entirely negative intervals. Posterior credible intervals associated to the random slope of Albania, France, Italy and Slovenia evidence a decreasing pattern of the industrial fishing while Algeria, Bosnia and Herzegovina and Turkey exhibit positive trends. Croatia, Montenegro, Morocco and Spain present a fairly stable dynamic.
	
	%\begin{figure}[H]
	%	\centering
	%    \caption{Artisanal and industrial fishing. Posterior 95 \% credible intervals from the joint model posterior sample of random effects by country.} \label{fig:random_effects_joint}
	%	\subfigure[Random intercepts. ]{\includegraphics[width=78mm]{images/intervals_intercept_joint.png}}
	%	\subfigure[Random slopes.]{\includegraphics[width=78mm]{images/intervals_trend_joint.png}}
	
	%\end{figure}
	
	\section{Discussion}
	We have presented a two-model approach to assess the temporal evolution of the European sardine landings in the Mediterranean countries from 1970 to 2014. We  first provided  a general overview on the fisheries, and secondly focused on a comparison   between industrial and artisanal fisheries. Both models dealt with general patterns  as well as individual characteristics of the countries in the study.  
	
	Overall results confirmed that Mediterranean fisheries were highly diverse from the beginning of the time series and this heterogeneity still remains over time. This general result is in agreement with a recent study that analyzed the convergence of the Mediterranean countries in terms of several ecological indicators (i.e. Marine Trophic Index, Fishing in Balance Index and Expansion Factor) during 1950-2010 \cite{pennino2017analysis}, showing strong temporal persistence in their fishery behaviours. The fact that
	Mediterranean fisheries are composed by a large number of small vessels operating in very local fisheries, with diverse cultural, social and economic characteristics, sharing the same resources, is recognised as the major challenge towards a sustainable management\cite{piroddi2015modelling}.
	
	At country level, interestingly, although Bosnia and Herzegovina is the country that reported the lowest landings and artisanal only, in the recent years has undergone expansion showing an increasing pattern in the sardine landings. 
	%Bosnia and Herzegovina is a small country with a very short coastline on the Mediterranean sea. Observed landings were the ones officially reported to FAO by the former Yugoslavia which were disaggregated to each member of the Yugoslavia. 
	However, this increasing trend could be due to the fact that from  1992 to 2010, when Bosnia and Herzegovina began to independently reports its catches to the FAO (Food and Agriculture Organization), the reconstructed domestic catches were 19 times larger than the data presented by FAO on behalf of Bosnia, with the difference being mainly due to unreported landings \cite{pauly2016global}.
	
	Turkey also showed a positive trend in its temporal dynamics.
	Turkish fishery expansion has been previously documented and it is probably a consequence of direct government support to the fishing sector as well as the implementation of technological advances. The modernization of small and large-scale fishing fleets (i.e. larger boats, of higher tonnage and engine horsepower, improved fishing gear, the use of high-technology equipment) is leading to the expansion of fishing in areas previously inaccessible to fishing vessels because of strong winds and in deep water areas \cite{harliouglu2011present}. 
	Sardine stock in the Aegean
	Sea is a shared stock between the Greek and the Turkish fishing fleets \cite{antonakakis2011assessment}.
	However, Greece, showed a quite stable dynamics, jointly with Croatia, Morocco and Spain.
	
	Albania, France, Italy, Montenegro and Slovenia showed a decreasing pattern in the European sardine landing temporal dynamic. This general decrease in landings is in agreement with other observations, such as a decreasing biomass and weakened biological state of the sardine stocks in the territorial waters of these countries \cite{brosset2017spatio}. 
	The decreasing trend might be explained by the combined effect of exploitation and environmental changes, especially in areas where sardine were clearly overexploited, such as the Adriatic Sea \cite{fao2014}. 
	
	\cite{tzanatos2014indications} showed that the European sardine landings went through a significant abrupt decline in the Western and Eastern Mediterranean sea in the mid-late 1990s, probably due to a parallel increase in the temperature regime. A catch-per-unit-of-effort proxy using data from the different Mediterranean Sea member states also showed significant correlation with temperature fluctuations and sardine landings, indicating the persistence of temperature influence on landings when the fishing effect is accounted for \cite{stergiou2016trends}. 
	However, although the Mediterranean Sea is strongly affected by large atmospheric transfers, forceful influences are also attributable to more local factors, such as river runoff, which induce contrasted conditions at small temporal and spatial scales. As a consequence, more local drivers must be investigated in order to better understand these fluctuations in these specific areas. 
	
	Industrial activity was bigger than the artisanal one in the 70s, at the beginning of the period of the study. However, the comparison of temporal dynamics showed by the Bayesian joint model highlights that even if industrial fishing started earlier, both fisheries then presented similar trends in the different countries. Indeed, our results highlighted a positive relationship between both types of fishing.
	The industrialization of fisheries, mainly through government subsides, without study and planning, is not sustainable in the long run, with direct consequences to fish stocks.

	For this specific case study, the use of the official landings may be a limitation as this data does not account for discards, by-catch and illegal, unreported and unregulated (IUU) catches. Despite these limitations, the application of Bayesian longitudinal mixed model provides a novel way to assess changes in fisheries exploitation by different countries at large scale. This study represents a new standpoint from which to explore species fisheries exploitation time series under a probabilistic framework.
	%Indeed, most of the sardine stocks are exploited by multiple fisheries and often by different member states. %Therefore, it is considered that for stocks shared both in terms of different countries and fleets exploiting them, a fishery management plan needs to include all fleets and countries exploiting the stock (STECF, 2012e). 
	
	\section*{Acknowledgements}\label{sec4}
	Gabriel Calvo's research was funded by the ONCE Foundation, the Universia Foundation and the Spanish Ministry of Education and Professional Training, grant FPU18/03101. Luigi Spezia's research was funded by the Scottish Government's Rural and Environment Science and Analytical Services Division. 
	Maria Grazia Pennino's research was funded by the Spanish Research project PELWEB (CTM2017-88939-R) funded by Spanish Ministry of Science, Innovation and Universities, the European Research Contract SPELMED (EASME/EMFF/2016/032) funded by EC EASME.
	We are grateful to Glenn Marion which comments improved the quality of the final paper.

	\bibliography{manuscript_bibliography}
	
\end{document}